\documentclass[aps,superscriptaddress,amsmath,twocolumn,amssymb,showpacs,
floatfix]{revtex4} \usepackage{epsfig}

\begin{document}

\title{Heterogeneity and growing lengthscales in the dynamics of
kinetically constrained lattice gases in two dimensions}

\author{Albert C. Pan}

\affiliation{Department of Chemistry, University of California,
Berkeley, CA 94720-1460}

\author{Juan P. Garrahan}

\affiliation{School of Physics and Astronomy, University of
Nottingham, Nottingham, NG7 2RD, UK}
 
\author{David Chandler}

\affiliation{Department of Chemistry, University of California,
Berkeley, CA 94720-1460}

\date{\today}

\begin{abstract}
We study dynamical heterogeneity and growing dynamical lengthscales in
two kinetically constrained models, namely, the one- and two-vacancy
assisted triangular lattice gases.  One of the models is a strong
glassformer and the other is a fragile glassformer.  Both exhibit
heterogeneous dynamics with broadly distributed timescales as seen in
the distribution of persistence times.  We show that the
Stokes-Einstein relation is violated, to a greater degree in the
fragile glassformer, and show how this violation is related to dynamic
heterogeneity.  We extract dynamical lengthscales from structure
factors of mobile particles and show, quantitatively, the growth of
this lengthscale as density increases.  We comment on how the scaling
of lengths and times in these models relates to that in facilitated
spin models of glasses.
\end{abstract}


\maketitle

\twocolumngrid

\section{Introduction}

The dramatic dynamical slowdown accompanying the formation of a glass
is a remarkable phenomenon
\cite{Ediger-et-al,Angell,Debenedetti-Stillinger}.  One explanation
for the underlying microscopic cause of this slowdown relies on the
presence of local steric constraints on the movement of particles
which make themselves felt to an increasing degree as the temperature
is lowered (or the concentration of particles is increased).
Kinetically constrained lattice gas models
\cite{Kob-Andersen,Jackle,Ritort-Sollich} are simple caricatures of
glassformers which employ local steric constraints as their sole means
to glassiness in the {\em absence} of any non-trivial static
correlations between particles (for alternative thermodynamic views of
the glass transition see e.g.  \cite{Tarjus-Kivelson,Xia-Wolynes}).
These constrained models have been extensively studied (see e.g.
\cite{Kob-Andersen,Jackle,Sellito-et-al,Toninelli-et-al,Marinari-Pitard}).
Our purpose here is to extend these studies to focus on the idea of
dynamical heterogeneity \cite{Ediger,Glotzer,Harrowell-et-al} as a
manifestation of excitation lines in space-time and to attempt
identification of scaling and universality classes in the dynamics of these
models
\cite{Garrahan-Chandler,Berthier-Garrahan,Jung-et-al,Whitelam-et-al,Berthier-et-al}.

The paper is organized as follows.  Section \ref{Models} describes the
two models we use as well as details of the computer simulations used
to study them.  Section \ref{dpt} looks at heterogeneous dynamics in
our models via the distribution of persistence times.  Section
\ref{slowdown} presents the scaling of the structural relaxation time
and diffusion constant, the implications of which lead to a discussion
of the breakdown of the Stokes-Einstein relation in section
\ref{SEBreakdown}.  Section \ref{dynlength} discusses the emergence of
a dynamical lengthscale and gives a quantitative characterization of
this length by analyzing structure factors of mobile particles.
Finally, we end with a discussion of our results in Section
\ref{Discussion}.

\section{Models and Computational Details}
\label{Models}

We present results for two kinetically constrained triangular lattice
gas (TLG) models introduced by J\"ackle and Kr\"onig \cite{Jackle}.
These two-dimensional models are variants of lattice models proposed
by Kob and Andersen \cite{Kob-Andersen}.  Each site of the triangular
lattice has six nearest neighbor sites and can hold at most one
particle.  A particle at site ${\bf i}$ is allowed to move to a
nearest neighbor site, ${\bf i}'$, if (i) ${\bf i}'$ is not occupied
and (ii) the two mutual nearest neighbor sites of ${\bf i}$ and ${\bf
i}'$ are also empty.  These rules coincide with a physical
interpretation of steric constraints on the movement of hard core
particles in a dense fluid \cite{Jackle}.  We call the model with
these rules the (2)-TLG because both mutual nearest neighbors need to
be empty in order to facilitate movement.  We also present results for
the (1)-TLG where the constraints are more relaxed: movement is
allowed as long as either of the mutual nearest neighbors is empty.
As with other kinetically constrained lattice gas models, the TLG has
no static interactions between particles other than those that
prohibit multiple occupancy of a single lattice site.  Therefore,
initial configurations can be generated by random occupation of empty
lattice sites by particles until the desired density is reached.

In the computer simulations, we investigated particle densities,
$\rho$, between 0.01 and 0.80 for the (2)-TLG and between 0.01 and
0.996 for the (1)-TLG.  The density $\rho$ = 1 corresponds to the
completely full lattice in both cases.  For the (2)-TLG, we used a
lattice with edge length $L$ = 128 for all densities.  There exists
the possibility in the (2)-TLG of initial configurations containing an
unmoveable structure which percolates throughout the system called a
backbone \cite{Kob-Andersen, Jackle}.  Since the dynamics obey
detailed balance, these backbones could never be destroyed in the
course of the simulation.  For the densities studied here, however,
$L$ = 128 is sufficiently large such that the probability of having
such a configuration is vanishingly small (see \cite{Jackle}).  For
the (1)-TLG, we used $L$ = 128 to $L$ = 2048.  A pair of vacancies in
the (1)-TLG can always diffuse freely \cite{Jackle} so this version of
the TLG does not suffer from the problem of backbones in the same way
as the (2)-TLG.  At higher densities, however, one still needs to
ensure that there are a sufficient number of potentially mobile
particles in the system such that the typical dynamics of the model
are observed.  For the (1)-TLG, the number of potentially mobile
particles at higher densities is approximately equivalent to the
number of vacancy pairs and therefore can be estimated as
$(1-\rho)^2L^2$.  The system sizes at the various densities for the
(1)-TLG simulations were chosen such that the number of potentially
mobile particles estimated in this way was always approximately 100.
For both models, periodic boundary conditions were used.

\begin{figure}[t]
\epsfig{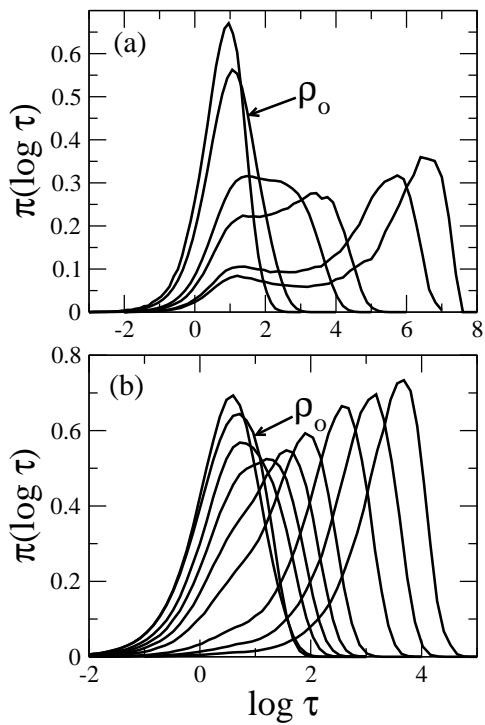}
\caption{\label{Dta} Distribution of local persistence times for (a)
the (2)-TLG model (from left to right) $\rho$ = 0.40, 0.50, 0.65,
0.70, 0.77 and 0.79; and (b) the (1)-TLG model (from left to right)
$\rho$ = 0.20, 0.60, 0.70, 0.75, 0.80, 0.85, 0.92, 0.95 and 0.97.  The
abscissae are given on a scale of logarithm base 10.}
\end{figure}
  
At each density, several hundred independent trajectories of lengths
10-100 times $\tau_{\alpha}$, where $\tau_{\alpha}$ is the time for
the self-intermediate scattering function at ${\bf q} = (\pi, 0)$
\cite{qnote} to reach $1/e$ of its initial value (see below), were
run.  Trajectories were stored logarithmically for later analysis
 (i.e. configurations were saved after 1, 2, 4, 8, 16, 32,
etc. sweeps).  At each state point, between 128 and 256 independent
trajectories were acquired.  Time was measured in Monte Carlo sweeps.
During each sweep, particles were chosen randomly and a move was
attempted.  For the higher density runs in both models, a continuous
time algorithm was used for greater efficiency \cite{Newman-Barkema}.
This algorithm involved making and updating a list of only those
particles which have the possibility of moving and choosing from among
those exclusively during every move.  The total time (in units of
Monte Carlo sweeps) was then updated accordingly by adding to it the
inverse of the number of mobile particles available during that
continuous time step.  The continuous time algorithm resulted in a
speed up of our simulations by as much as 1-2 orders of magnitude for
the highest density runs in both the (1)-TLG and the (2)-TLG.
Finally, for the distribution of site persistence times (see below),
statistics were gathered over runs of very large systems ($L = 1024$
and $2048$).

\begin{figure}[b]
\epsfig{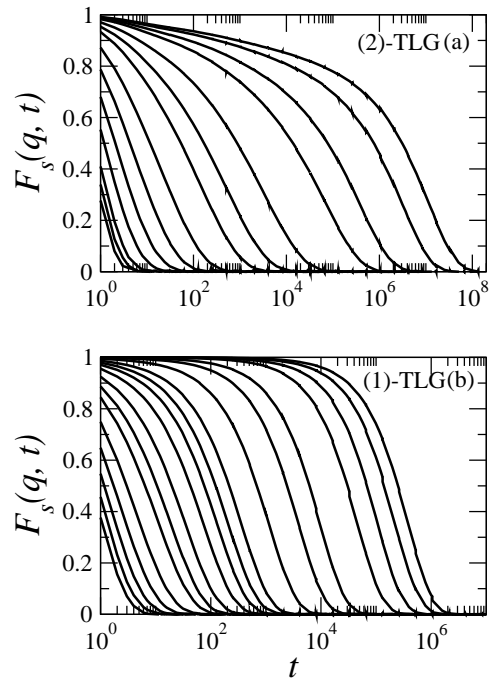}
\caption{\label{fskpi} Self-intermediate scattering function at
wavevector ${\bf q} = (\pi, 0)$ for (a) the (2)-TLG (from left to right)
$\rho$ = 0.01, 0.05, 0.10, 0.20, 0.30, 0.40, 0.50, 0.60, 0.65, 0.70,
0.75, 0.77, 0.79, 0.80; and (b) the (1)-TLG (from left to right)
$\rho$ = 0.20, 0.30, 0.40, 0.50, 0.60, 0.70, 0.75, 0.80, 0.85, 0.88,
0.90, 0.92, 0.95, 0.97, 0.98, 0.99, 0.993, 0.995, 0.996. The unit of
length is the lattice spacing. }
\end{figure}

\section{Heterogeneous dynamics and the Distribution of persistence times}
\label{dpt}

\begin{figure*}[t]
\epsfig{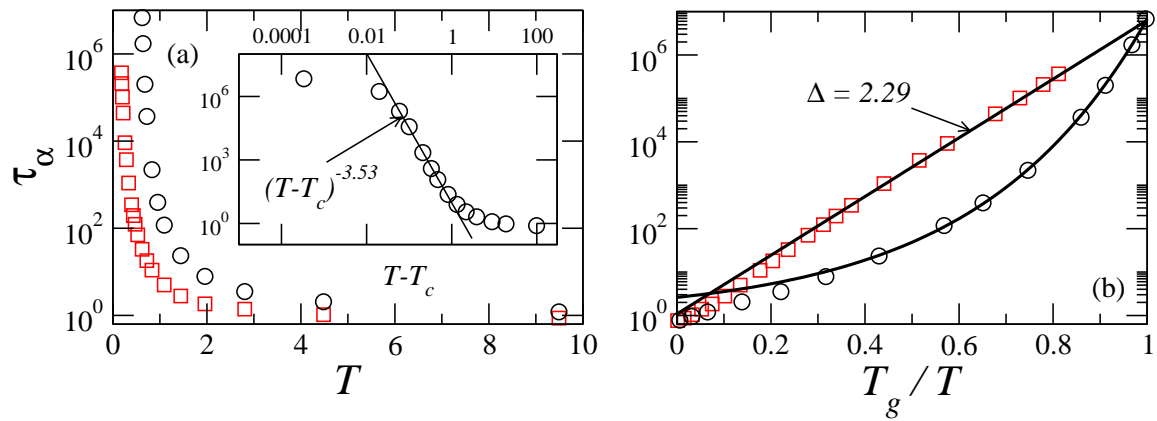}
\caption{\label{ang} (a) Structural relaxation time as a function of
temperature.  Circles and squares correspond to the (2)-TLG and
(1)-TLG, respectively, throughout the paper.  The inset shows the same
data for the (2)-TLG plotted versus $T - T_c$ and a power law fit
(solid line) to a portion of the data.  (b) Structural relaxation as a
function of scaled reciprocal temperature.  Here $T_g$ is such that
$\tau_\alpha(T_g) = 10^7$.  The lines are fits to the relaxation times
at low temperatures.  For the (1)-TLG we use the Arrhenius form
$\ln{\tau_\alpha} \propto \Delta / T$, with $\Delta \approx 2.29$.
For the (2)-TLG we use a double exponential \cite{Toninelli-et-al},
$\ln{\tau_\alpha} \propto \exp{(a / T)}$, with $a \approx 1.76$. }
\end{figure*}

\begin{figure}[t]
\epsfig{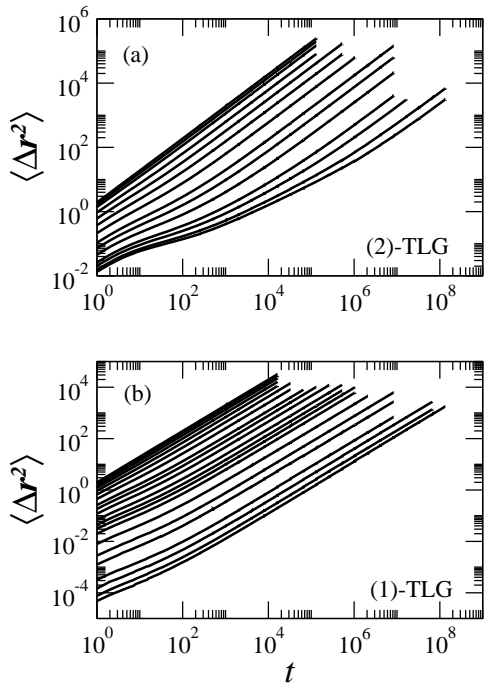}
\caption{\label{Dmsd} The mean-squared displacement: (a) the
(2)-TLG, $\rho = 0.01$ to $0.80$; (b) the (1)-TLG, $\rho = 0.01$ to
$0.996$.}
\end{figure}

A central phenomenon behind our perspective of glasses is dynamical
heterogeneity \cite{Ediger,Glotzer}.  A direct measure of
heterogeneous dynamics in glassy systems is the idea of persistence
times \cite{Berthier-Garrahan, Jung-et-al}.  Fig.\ \ref{Dta}(a) and
Fig.\ \ref{Dta}(b) show the distribution of site persistence times,
$\pi(\log \tau)$, in the (2)-TLG and (1)-TLG at various densities;
that is, the distribution of times, given an initial configuration of
the lattice, before the first change at a particular site occurs,
either due to an empty site being filled or a filled site becoming
empty.  These distributions are multi-point functions because they
depend not only on the state of a lattice site at the initial time and
the time when it changes, but also on all intervening points in time.
As has been shown in spin-facilitated models \cite{Berthier-Garrahan},
three distinct dynamical regimes are observed: (i) at low densities,
there is a single peak at small relaxation times indicating
homogeneous fast dynamics; (ii) at intermediate densities, two peaks
develop and co-exist, one at faster times and the other at slower
times, indicating heterogeneous, fluctuation dominated, dynamics;
(iii) as density is increased even higher, the peak at faster times
becomes suppressed relative to the peak at slower times and the
dynamics again become homogeneous and slow.  In region (ii), or the
crossover region, the dynamics are broadly distributed over several
orders of magnitude in time.  Following \cite{Berthier-Garrahan}, 
we can define, qualitatively, an onset
density, $\rho_o$, for both models where the dynamics begin to feel
the influence of dynamical heterogeneity and thereby lose their
mean-field character, as well as a crossover density $\rho_c$ where
slow processes begin to dominate.  For the (2)-TLG, $\rho_o = 0.50$
and (we anticipate) $\rho_c$ = 0.80 and for the (1)-TLG, $\rho_o =
0.60$ and $\rho_c$ = 0.85.

In the next section, we turn to two two-point functions which have
been the more conventional measures of glassy dynamics: the
self-intermediate scattering function and the mean-squared
displacement.

\section{Dynamical slowdown}
\label{slowdown}

\begin{figure}
\epsfig{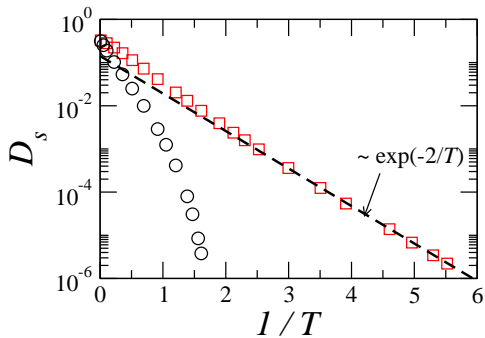}
\caption{\label{Dmsd-scal} Self-diffusion constant as a function of
inverse temperature.}
\end{figure}

A basic ingredient of a glassformer is a precipitous dynamical
slowdown over a narrow range of temperatures or densities.  This
characteristic can already be seen qualitatively in the distribution
of persistence times by looking at the movement of the mean of the
distributions as $\rho$ increases.  For example, in the (2)-TLG, from
$\rho$ = 0.70 to $\rho$ = 0.77, the mean persistence time increase 2-3
orders of magnitude.  The traditional measure of this slowdown is the
self-intermediate scattering function, $F_s(q, t)$ = $\langle e^{i{\bf
q}\cdot({\bf r}_i(t) - {\bf r}_i(0))}\rangle$, particularly its decay
at wavevector ${\bf q} = (\pi, 0)$, Fig.\ \ref{fskpi}(a) and Fig.\
\ref{fskpi}(b).  Here, ${\bf r}_i(t)$ denotes the position of particle
$i$ at time $t$.  The angled brackets, $\langle\cdots\rangle$, denote
an average over different pairs of configurations along a trajectory
separated by a given time interval.  The decay of the scattering
function to $1/e$ at this wavevector is typically defined to be
$\tau_{\alpha}$, or the structural relaxation time, as it gives a
sense of how density fluctuations relax at relatively short
lengthscales.

Experiments report the scaling of viscosity versus inverse temperature
\cite{Angell}.  Therefore, for the particular case of a kinetic
lattice gas, one would like to make connections between structural
relaxation time and viscosity, and density and inverse temperature.
In both experiments and computer simulations
\cite{RichterFrickFarago,OnukiPRE}, it has been shown that the
structural relaxation time scales like the viscosity.  If one imagines
that the concentration of vacancies, $c$, in the TLG models are like
excitations or fluctuating regions of high energy, then it is
reasonable to define a reduced temperature, $T$, such that $c$
$\equiv$ $(1-\rho)$ $\equiv$ $e^{-1/T}$ or:
\begin{equation}
-\ln(1-\rho)\equiv 1/T.  
\end{equation}
The structural relaxation time plotted versus $T$ and $1/T$ is
shown in Fig.\ \ref{ang}(a) and Fig.\ \ref{ang}(b), respectively.
The inset of Fig.\ \ref{ang}(a) shows a plot of $\tau_{\alpha}$
versus $T-T_c$ for the (2)-TLG where $T_c$ is taken to be the
temperature at the crossover density, $\rho_c$ (as defined in section
\ref{dpt}), and the black line is a power law fit to a portion of the
data in the manner of mode coupling theory (MCT) \cite{mct}.  In MCT,
$T_c$ is a critical temperature where time scales diverge.  We see
that the MCT fit is valid for almost five orders of magnitude in time
even though there is no dynamical arrest at $T_c$, as the relaxation
time of the (2)-TLG diverges only at $\rho=1$ \cite{Toninelli-et-al}.
A similar MCT power law fit (not shown) can be made for the (1)-TLG,
valid for about 2 orders of magnitude.

Fig.\ \ref{ang}(b) shows the relaxation time on a reduced temperature
scale in the manner proposed by Angell \cite{Angell}.  Here, $T_g$ is
defined as the temperature at which $\tau_{\alpha} = 10^7$.  We see
that the relaxation time of the (1)-TLG is Arrhenius growing as
$\tau_{\alpha} \sim c^{-\Delta}$ with $\Delta = 2.29$, as $\rho
\rightarrow 1$.  The relaxation of the (2)-TLG in contrast is
super-Arrhenius, and the low temperature data can be fit with a double
exponential in $1/T$ \cite{Toninelli-et-al}.  In this context, the
(1)-TLG is a strong glassformer whereas the (2)-TLG is a fragile one.

\begin{figure}
\epsfig{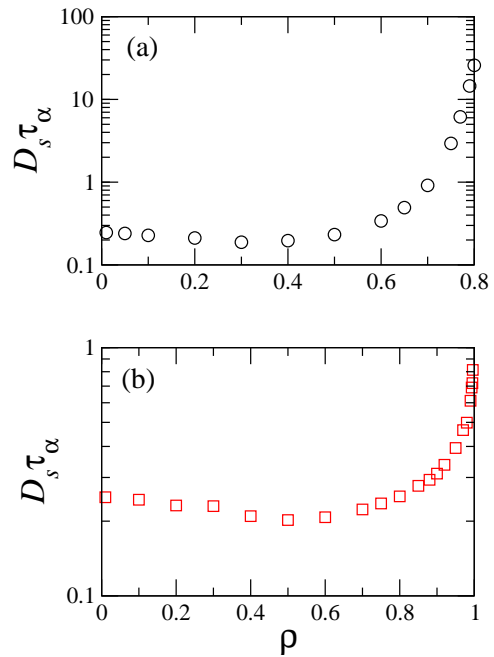}
\caption{\label{SE} Stokes-Einstein violation in the (a) (2)-TLG
and (b) (1)-TLG.  }
\end{figure}

\begin{figure}
\epsfig{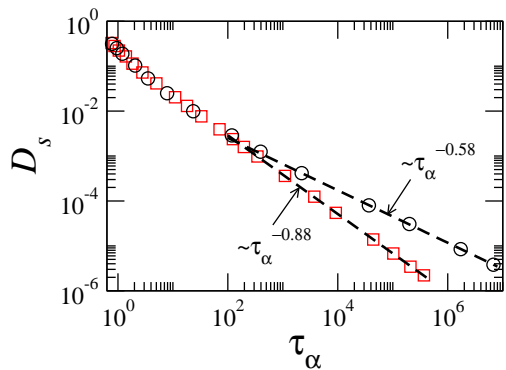}
\caption{\label{fSE} Fractional Stokes-Einstein exponent: scaling of
  the self-diffusion constant with relaxation time.  The dashed lines
  are fits to the data at longer times.}
\end{figure}

Interestingly, the time exponent $\Delta \approx 2.3$ of the (1)-TLG
is the same as that for the one-spin facilitated Fredrickson-Andersen
(FA) model in dimension $d=2$, obtained in renormalization group (RG)
calculations and observed numerically \cite{Whitelam-et-al}.  This
suggests that the (1)-TLG may be in the universality class of the FA
model, the prototypical kinetically constrained spin model for a
strong glassformer \cite{Trivial}.  We will examine more of these
dynamic scaling relations below.

\begin{figure*}
\epsfig{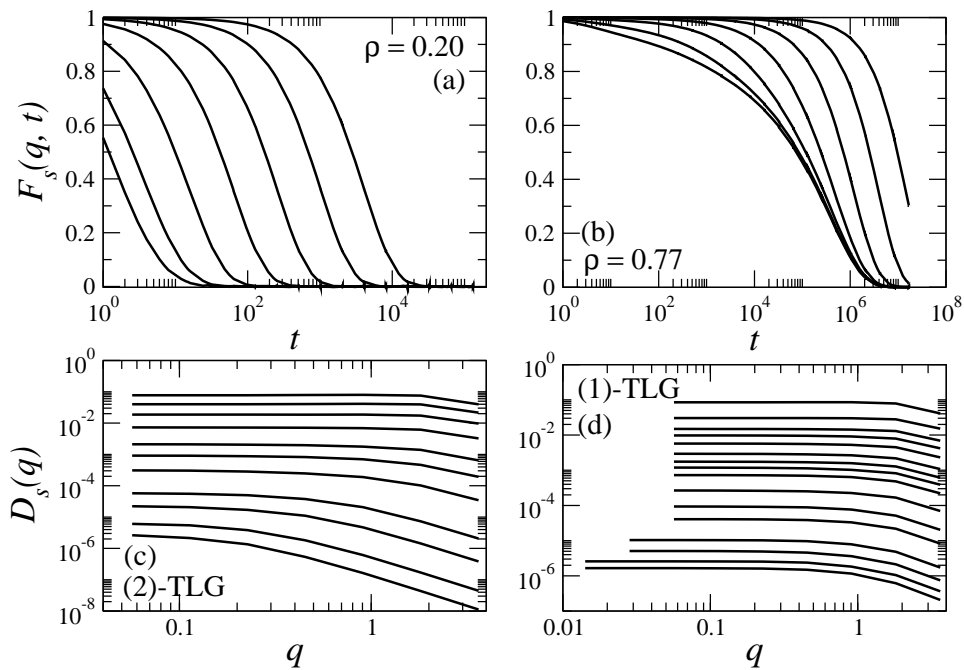}
\caption{\label{diffq} (Top) Self-intermediate scattering function for
the (2)-TLG at various wavevectors: (a) $\rho = 0.20$, (b) $\rho =
0.77$.  For both graphs, from left to right: $q$ = $(\pi, 0)$,
$(\pi/2,0)$, $(\pi/4, 0)$, $(\pi/8, 0)$, $(\pi/16, 0)$, $(\pi/32, 0)$,
$(\pi/64, 0)$.  (Bottom) $D_s(q) \equiv 1/\tau(q)q^2$ as a function of
$q$ at various densities: (c) (2)-TLG, (from top to bottom) $\rho$ =
0.20 to 0.80; (d) (1)-TLG (from top to bottom), $\rho$ = 0.40 to
0.996.  The higher density curves for the (1)-TLG include smaller $q$
values because of larger system sizes (see section II).}
\end{figure*}

The mean-squared displacement, $\langle|\Delta {\bf r}_i(t)|^2
\rangle$ = $\langle|{\bf r}_i(t) - {\bf r}_i(0)|^2\rangle$, is shown
in Fig.\ \ref{Dmsd}.  The self-diffusion coefficient, $D_s$, is
defined as $D_s = \lim_{t\rightarrow\infty}\langle|\Delta {\bf
r}_i(t)|^2\rangle/4t$.  We see that at low
densities, $D_s$ for the (1)-TLG and (2)-TLG coincide.  
At higher densities, $D_s$ for the (1)-TLG is
Arrhenius, and scales as $D_s \sim c^2$ as $\rho \rightarrow 1$, see
Fig.\ \ref{Dmsd-scal}.  This result is in agreement with the
analytical prediction of \cite{Toninelli-et-al}.  Moreover, this
is also the scaling of the diffusion constant for a probe molecule
coupled to the FA model in any dimension \cite{Jung-et-al}, further
evidence that the (1)-TLG is in the FA model universality class.  On
the other hand, $D_s$ for the (2)-TLG is super-Arrhenius (see Fig.\
\ref{Dmsd-scal}).  The behavior of $D_s$ is similar, qualitatively, to
that of $\tau_{\alpha}$.  The quantitative difference in their scaling
with density, however, is significant, and is an indication that
relaxation behaviors at short and long lengthscales are not the same.
We turn to this issue now in both the (1)-TLG and the (2)-TLG.

\begin{figure}[b]
\epsfig{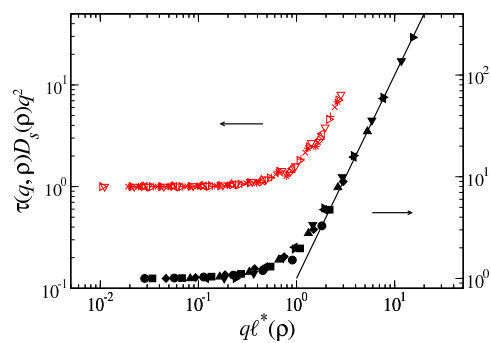}
\caption{\label{scaling} Data from Fig.\ \ref{diffq} collapsed onto a
  master curve. The filled symbols correspond to the (2)-TLG and the
  open symbols correspond to the (1)-TLG.  The straight line is $q^2$.}
\end{figure}

\section{The breakdown of the Stokes-Einstein Relation}
\label{SEBreakdown}

An important ramification of broadly distributed heterogeneous
dynamics is the breakdown of mean-field dynamical relations such as
the much studied Stokes-Einstein (SE) relation.  This relation says
that diffusion scales inversely with the relaxation time,
$D_s\tau_{\alpha} \sim \mbox{constant}$.  It is a quantitative
statement of the expectation that the dynamical behavior of normal
liquids should be similar at all but the smallest lengthscales.  In supercooled liquids
this simple mean-field approximation fails
\cite{Chang-Sillescu,Swallen-et-al}, and, given the results of
previous sections, we would expect a similar violation of the SE
relation in the TLG models.  We see from Fig.\ \ref{SE} that the SE
relation is indeed violated for both the (2)-TLG and the (1)-TLG, the
effect being more pronounced in the fragile case.  Moreover, we see
that the densities at which the product $D_s\tau_{\alpha}$ begins
deviating from constancy coincide with the onset densities,
$\rho_o$'s, extracted from the distribution of persistence times.
This observation reinforces the idea that it is the fluctuation
dominated nature of the dynamics that leads to the SE breakdown
\cite{Jung-et-al}.

SE violation implies that the self-diffusion constant does not scale
with the structural relaxation time as $\tau_{\alpha}^{-1}$.  One
possibility is that it obeys a fractional SE law, $D_s \sim
\tau_{\alpha}^{-\xi}$ where $\xi < 1$.  This is observed in
experiments \cite{Swallen-et-al}, and is obtained theoretically for
probe diffusion in the FA and East models
\cite{Jung-et-al} (see also \cite{Schweizer-Saltzman}).  Figure
\ref{fSE} shows that the diffusion constant also
obeys a fractional SE law in the TLG models.  The SE exponent is $\xi \approx 0.88$ for
the (1)-TLG, which is the value expected for the FA model in $d=2$,
$\xi \approx 2/2.3$ \cite{Jung-et-al}.  In the case of the (2)-TLG,
despite the fact that $D_s$ and $\tau_{\alpha}$ are both
super-Arrhenius, we find that the scaling exponent is temperature
independent at large densities, $\xi \approx 0.58$.  The deviation of
this exponent from 1 is
larger than that for both the FA and East models in two dimensions
\cite{Jung-et-al}.  It indicates a larger violation of the SE law,
consistent with the fact that the (2)-TLG is more fragile than either
of those models \cite{Toninelli-et-al}.

\section{Dynamical Lengthscales}  
\label{dynlength}

\subsection{Indication of a dynamical lengthscale from a two-point function}

\begin{figure}[t]
\resizebox{\columnwidth}{!}{
\includegraphics{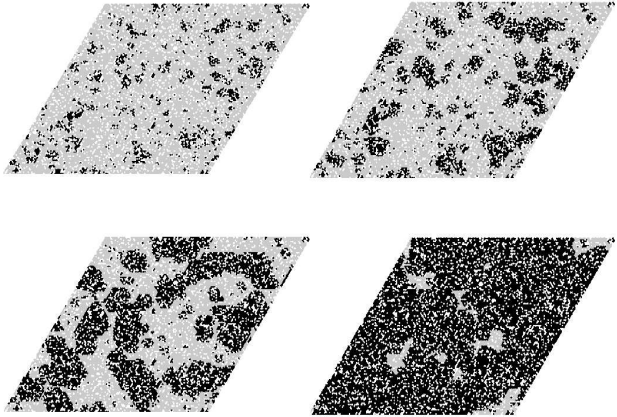}}
\caption{Growth of mobile particle regions as a function of
  observation time $\Delta t$ at $\rho = 0.77$ in the (2)-TLG.  Black
  and grey regions indicate the location of particles and white
  regions indicate empty lattice sites.  Particles colored
  in black have moved at least one lattice spacing in a time $\Delta
  t$ whereas particles colored in gray have not.  (Top, from left to right:
  $\Delta t = 10^3$ and $10^4$; bottom, from left to right: $\Delta t = 10^5$
  and $10^6$; $\tau_{\alpha} \sim 10^5$ at this density).}
\label{2tlg}
\end{figure}

Since the growth in timescales and the violation of the
Stokes-Einstein relation in the TLG models are clearly not tied to a growth in static
lengthscales, we turn now to the discussion of dynamical lengthscales.
Such a lengthscale can be inferred from examining the relaxation
behavior of the self-intermediate scattering function over more than
one wavevector at different densities.  One can appreciate this fact
qualitatively by looking at the decay of $F_s(q, t)$ for the (2)-TLG
over several values of $q$ at low and high density, Fig.\ \ref{diffq}
(top), \cite{Berthier-et-al}.  At low density, the decay of the
various curves looks similar at all wavevectors (except for the
largest wavevector) whereas at high density, even the curves at
intermediate wavevector differ greatly from the simple exponential
form seen at smaller wavevectors.  The high density curves bunch up at
intermediate to large ${\bf q}$ indicating that the relaxation
behavior at these lengthscales is different (i.e. slower) than one
would expect from the behavior at larger lengthscales.  Similar
behavior has also been observed in the Kob-Andersen kinetic lattice
gas model and kinetically constrained spin models
\cite{Kob-Andersen,Berthier-et-al}.

\begin{figure}[t]
\resizebox{\columnwidth}{!}{
\includegraphics{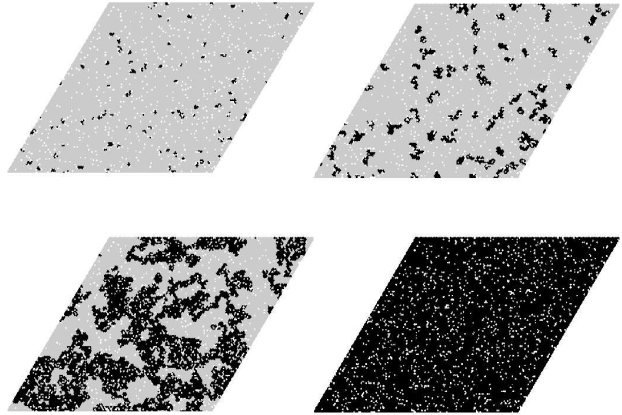}}
\caption{Same as Fig.\ \ref{2tlg} for the (1)-TLG.  Here, $\rho$ =
0.95.  (Top, from left to right: $\Delta t = 10$ and $10^2$; bottom,
from left to right: $\Delta t = 10^3$ and $10^4$; $\tau_{\alpha} \sim
10^3$ at this density).}
\label{1tlg}
\end{figure}

To quantify the above behavior, we proceed as in
\cite{Berthier,Ediger-et-al-2}.  In the hydrodynamic regime, we have
$\lim_{q \rightarrow 0} F_s(q, t) \sim \exp(-D_s q^2 t)$, and one
expects the product $D_s \tau(q) q^2$ to be independent of ${\bf q}$, where
$\tau(q)$ is the time when the intermediate scattering
function at wavevector $\bf{q}$ decays to $1/e$.  In Fig.\ \ref{diffq} (bottom), we
plot the quantity $D_s(q) \equiv 1/\tau(q) q^2$ as a function of $q$ at
various densities.  A flat line independent of $q$ indicates normal
diffusive behavior whereas a downward bend signifies a change to
sub-diffusive behavior.  As density increases, the curves begin to
bend at a smaller and smaller wavevector, $q^{*}$.  This behavior is
indicative of a growing dynamical lengthscale as density is increased
\cite{Berthier,Ediger-et-al-2,Berthier-et-al}.

Following the prescription of \cite{Berthier-et-al}, we extract a
lengthscale, $\ell^*$, from Fig.\ \ref{diffq} as $\ell^* \sim
\sqrt{D_s(q \rightarrow 0) \tau_{\alpha}}$.  This lengthscale determines
the onset of Fickian diffusion.  Using $\ell^*$ to rescale
space, and using $\tau(q)q^2$ to rescale time, the data from Fig.\
\ref{diffq} can be collapsed onto a master curve, Fig.\ \ref{scaling}.

\subsection{Direct observation and quantification of a dynamical
heterogeneity lengthscale}

\begin{figure*}[t]
\epsfig{file = 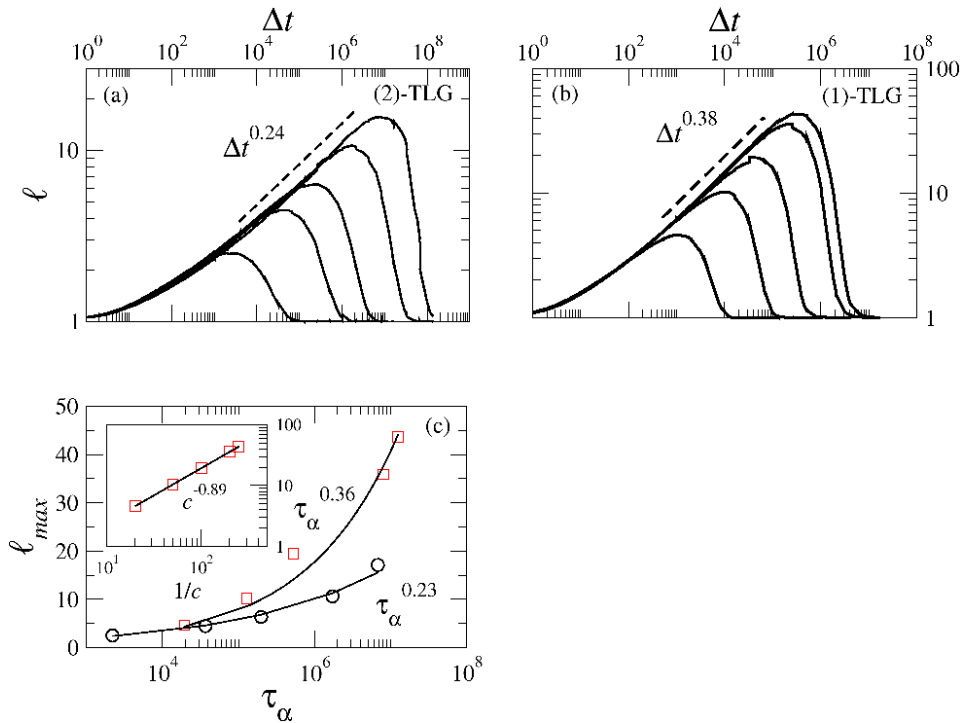, width=5in, clip}
\caption{\label{xi} (Top) Growth of dynamical heterogeneity length as
a function of observation time, $\Delta t$, for (a) the (2)-TLG,
$\rho$ = 0.70 to 0.80; and (b) the (1)-TLG, $\rho$ = 0.95 to 0.996.
(Bottom) (c) Scaling of the maximum value of the dynamical heterogeneity
length, $\ell_{max}$, from (a) and (b) with relaxation time.  
The inset shows $\ell_{max}$ versus $c$ for the (1)-TLG, on a
log-log scale to highlight power law scaling.}
\end{figure*}

We can study dynamical lengthscales in the TLG models directly by
observing a trajectory over a time $\Delta t$ and coloring particles
which have moved at least one lattice spacing.  Snapshots of applying
this procedure to trajectories of the (2)-TLG and the (1)-TLG at high
particle densities over progressively longer $\Delta t$'s are shown in
Fig.\ \ref{2tlg} and Fig.\ \ref{1tlg}.  Mobility is indeed correlated:
mobile particles are clustered and the clusters of mobility at earlier
$\Delta t$ act as seed particles from which subsequent mobility grows.
Moreover, there is a qualitative difference in the shape of the
clusters in the (2)-TLG and the (1)-TLG.  In the fragile model, the
clusters are smooth and more anisotropic, indicating a directed growth
of mobile regions.  In the strong model, the clusters are rugged and
isotropic.  These observations can be understood as arising from the
difference in the local constraints of both models.  The strict two
site facilitation rule of the (2)-TLG requires cooperative,
hierarchical rearrangement of particles for movement whereas the one
site facilitation rule of the (1)-TLG allows for the random diffusion
of vacancy pairs \cite{Jackle}.  These same correlations between
fragility and the smoothness of interfaces and between slow and fast
dynamically heterogeneous regions are present in other facilitated
models \cite{Garrahan-Chandler,Whitelam-et-al,NEF}.

To quantify the above ideas, we extract a dynamical lengthscale,
$\ell(\Delta t)$, from structure factors of the mobile particles
\cite{Glotzer,Garrahan-Chandler,Lacevic-et-al,Berthier}.  Motivated by
the mobility criterion described in the previous paragraph and
depicted in Fig.\ \ref{2tlg} and Fig.\ \ref{1tlg}, we consider the binary field $n_{\bf
r}(t; \Delta t) = p_{\bf r}(t)[1-p_{\bf r}(t+\Delta t)]$, which gives
a signal only when there is particle motion at ${\bf r}$ over the
range of time between $t$ and $t+\Delta t$.  Its structure factor is a
four point function: it measures a correlation function which depends
on two points in time, $t$ and $t + \Delta t$, and two points in
space, ${\bf r}$ and ${\bf r'}$.

We define the structure factor for the mobility field as the following
normalized correlation function:
\begin{equation}
\label{skeq}
S({\bf q}; \Delta t) = \frac{1}{L^2}\frac{\langle \delta n_{\bf q}(t;
  \Delta t) \delta n_{-\bf q}(t; \Delta t)\rangle}{\langle\delta
  n_{\bf r}(t; \Delta t)^2\rangle}
\end{equation} 
where $\delta n_{\bf r}(t; \Delta t)$ = $n_{\bf r}(t; \Delta t) -
\langle n_{\bf r}(t; \Delta t)\rangle$ is the deviation of $n_{\bf r}$
from its average value and $\delta n_{\bf q}(t; \Delta t)$ is the
Fourier transform of $\delta n_{\bf r}(t; \Delta t)$:
\begin{equation}
\delta n_{\bf q}(t; \Delta t) = \sum_{\bf r} \exp\left(\frac{2\pi
    i}{L^2}{\bf r} \cdot {\bf q}\right)\delta n_{\bf r}(t; \Delta t).
\end{equation} 
The angled brackets, $\langle\cdots\rangle$, denote an average over
different pairs of configurations along a trajectory separated by a
given time interval $\Delta t$.  We then define the lengthscale,
$\ell(\Delta t)$, to be proportional to the inverse of the first
moment, $\bar{q}_{\Delta t}$, of the circularly averaged structure
factor, $\tilde{S}(q_n; \Delta t)$ \cite{Amar-et-al}.  That is,
$\ell(\Delta t) \sim 1/ \bar{q}_{\Delta t} $ where:
\begin{equation}
\bar{q}_{\Delta t} = \sum_{n}q_n\tilde{S}(q_n; \Delta
t)/\sum_{n}\tilde{S}(q_n; \Delta t).
\end{equation}
Here, $q_n$ = $2\pi n/L$ and $n$ = 0, 1, 2, ..., $L/2$.  The
lengthscale was normalized such that $\ell$ extracted from the
structure factor of a random configuration of particles on the lattice
(i.e. an ideal gas) was unity.

Fig.\ \ref{xi}(a) and Fig.\ \ref{xi}(b) show $\ell(\Delta t)$ at
various densities for the (2)-TLG and the (1)-TLG.  The basic shape of
these curves is as expected: as $\Delta t \rightarrow 0$, mobility is
sparse and uncorrelated so $\ell$ approaches unity and as $\Delta t
\rightarrow \infty$, everything becomes mobile and $\ell$ once again
tends towards unity.  In between, as the pictures in Fig.\ \ref{2tlg}
and Fig.\ \ref{1tlg} suggest, mobility clusters together and grows.
Looking at the maximum of these different curves, $\ell_{max}(\Delta
t_{max})$, a growing lengthscale is clearly evident as $\rho$
increases.  It is important to note that, in general, $\ell_{max} \neq
\ell^*$ \cite{Berthier-et-al}.
   
Fig.\ \ref{xi}(c) shows the value of $\ell_{max}$ plotted versus the
structural relaxation time for both the (2)-TLG and the (1)-TLG.  At
short relaxation times, 
the curves merge and approach the ideal gas value of one.
 As relaxation times increases, the dynamical heterogeneity
lengthscale for the strong version of the model is always larger than that of
the fragile version at a
fixed value of $\tau_{\alpha}$ \cite{Garrahan-Chandler}.  We also find that the
observation time, $\Delta t_{max}$, at which the maximum lengthscale,
$\ell_{max}$, occurs, scales with the structural relaxation time,
$\tau_{\alpha}$, for both models (not shown).  

If the (1)-TLG is in the universality class of the FA model, then we
would expect $\ell_{max}$ to scale as a power of both the excitation
concentration, $\ell_{max} \sim c^{-\nu}$, and of the relaxation time,
$\ell_{max} \sim \tau_{\alpha}^{1/z}$.  This appears to be the case,
as shown in Fig.\ \ref{xi}(c).  For the correlation and dynamic
exponents we find $\nu \approx 0.89$ and $1/z \approx 0.36$, in
reasonable agreement with Ref.\ \cite{Whitelam-et-al}, $\nu \approx
0.7$ and $1/z = \nu/\Delta \approx 0.3$ \cite{Trivial2}.  The $z$ exponents shown
in Fig.\ \ref{xi}(c) are what we would expect from Fig.\ \ref{xi}(b)
where a range of $\ell(\Delta t)$ curves at different densities 
merge at early times and display power law scaling with
similar exponents. 

As alluded to earlier, the dynamics of the (1)-TLG at high densities
is controlled by the motion of vacancy pairs.  The physics of these
vacancy pairs is similar to excitations in the FA model.  Vacancy
pairs have the ability to interact with other lattice vacancies in
order to branch and coalesce.  It is important to note that evidence
of these interactions can only be seen in simulations of large enough
system sizes where the number of vacancy pairs is approximately
50-100.  As mentioned in section \ref{Models}, this requirement leads
to system sizes, for example, of $L=2048$ for $\rho = 0.995$.

\begin{figure}[t]
\epsfig{file = 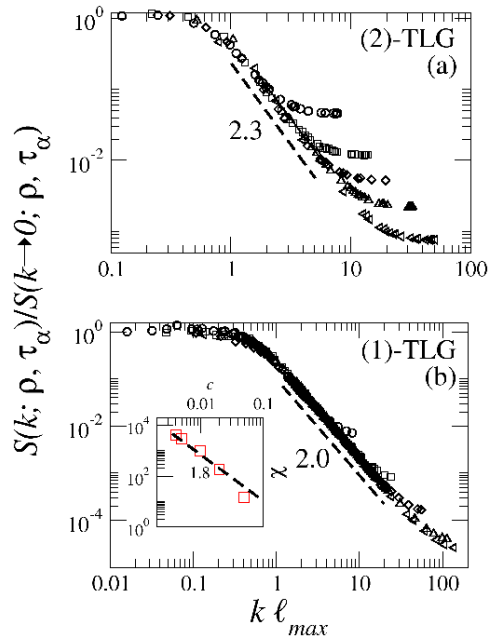, width=2.5in, clip}
\caption{Structure factors of the mobility field, equation
  (\ref{skeq}), measured at $\Delta t = \tau_{\alpha}$, 
for various densities in the (a) (2)-TLG and the 
(b) (1)-TLG.  The axes are scaled as indicated.}
\label{skscaled}
\end{figure}

Fig.\ \ref{skscaled} shows the structure factors of the mobility
  field, equation (\ref{skeq}), measured at the structural relaxation
  time, $\Delta t = \tau_{\alpha}$ for the (2)-TLG and the (1)-TLG.
  The curves are scaled in a manner suggestive of a coarsening
  process.  The collapse of the various structure factors
  \cite{largek} implies that, in the glassy regime, increasing density
  corresponds to a coarsening of dynamic heterogeneity fields in
  space-time.  An appreciation for the self-similarity of dynamic
  heterogeneity fields at different temperatures has already been
  noted in spin-facilitated models \cite{Garrahan-Chandler,
  Whitelam-et-al}.

Both sets of scaled structure factors in Fig.\ \ref{skscaled} have an
intermediate power law regime going as $k^{-2.3}$ for the (2)-TLG and
$k^{-2}$ for the (1)-TLG.  One explanation for the difference
in exponents could be the following.
Porod law scaling, $k^{-(d+1)}$, arises from a
system which is extensive in interfaces.  In two dimensions, this
implies that a system with a scaling exponent closer to 3 would have
smoother interfaces.  This interpretation is consistent with the
snapshots of the 
dynamic heterogeneity fields in Fig.\ \ref{2tlg} and Fig.\ \ref{1tlg}.

The structure factors for the (1)-TLG imply a value of the dynamical
exponent
$\eta$ very close to zero, $S(k) \sim 1/k^{2-\eta}$ as
$k\rightarrow\infty$.  
This gives a prediction for the exponent
$\gamma$ via the scaling relation $\gamma = (2-\eta)\nu_{\perp}$ of
$\gamma \approx 1.8$.  The $\gamma$ exponent controls the scaling of
the dynamic susceptibility, $\chi \equiv S(q\rightarrow 0; \rho,
\tau_{\alpha})$ with the concentration of excitations. 
The inset to Fig.\ \ref{skscaled}(b) shows that 
this expectation is approximately satisfied \cite{Whitelam-et-al}.     

\section{Discussion}
\label{Discussion}

Despite their simplicity, the constrained lattice gas models we have
studied show the essential features of glass forming liquids, such as
a precipitous dynamical slowdown and dynamical heterogeneity.  They
are intermediate between the fully coarse grained kinetically
constrained spin models such as the FA and East models, and atomistic
models such as the binary Lennard-Jones mixture.  We find a broad
distribution of persistence times, especially in the (2)-TLG
(Fig.\ \ref{Dta}).  
From the scaling of
the structural relaxation time it follows that the (2)-TLG is a
fragile model and the (1)-TLG is a strong one, consistent with the
predicted scaling of the self-diffusion constant in \cite{Toninelli-et-al}.  
Fragile behavior
versus non-fragile behavior coincides with hierarchical versus
diffusive propagation of excitations \cite{Jackle,Palmer-et-al}, and
the former follows from directional persistence
\cite{Garrahan-Chandler} as evident from the patterns of dynamic
heterogeneity seen in Fig.\ \ref{2tlg} and Fig.\ \ref{1tlg}.  
Dynamic heterogeneity produces length-time
scaling and decoupling phenomena.  Dynamic heterogeneity is present in
both strong and fragile materials, not only in the latter.  This is
consistent with recent molecular dynamics simulation on silica
\cite{Vogel-Glotzer} and earlier theoretical  predictions
\cite{Garrahan-Chandler,Berthier-Garrahan}.

\acknowledgments 

We are grateful to Robert Jack for important discussions.  This work
was supported by the US National Science Foundation, by the US
Department of Energy grant no.\ DE-FE-FG03-87ER13793, by EPSRC grants
no.\ GR/R83712/01 and GR/S54074/01, and University of Nottingham grant
no.\ FEF 3024.  A.C.P. is an NSF graduate research fellow.  This
research used resources of the National Energy Research Scientific
Computing Center, which is supported by the Office of Science of the
U.S. Department of Energy under Contract No. DE-AC03-76SF00098.


\begin{thebibliography}{99}
\bibitem{qnote} The wavevector dependent quantity, $F_s(q, t)$, was
calculated as a Fourier transform on the square lattice. A
correspondence can be made between the reciprocal lattice vectors of
the square and triangular lattices as in \cite{Jackle}.
\bibitem{Trivial} An alternative explanation is that relaxation in the
  (1)-TLG can simply be accounted for by the free diffusion of vacancy
  pairs, with interactions playing no role asymptotically.  In that
  case $\tau_\alpha$ would scale as $c^{-2}$ and the deviations
  measured numerically would correspond to logarithmic corrections
  characteristic of two-dimensional diffusion.  This is the situation,
  for example, in the square-plaquette spin model \cite{PQ}.
\bibitem{Trivial2} If interactions play no role in the (1)-TLG, then
  we would expect to observe $\nu=1$ and $z=2$.
\bibitem{largek} The discrepancy at large $k$ is due to probing the
  granularity of the lattice.
\end{thebibliography}
\end{document}